\documentclass[]{article}

\usepackage[a4paper, left=0.6in, right=0.6in, top=0.6in, bottom=0.6in, footskip=.25in]{geometry}
\usepackage{authblk}
\usepackage{graphicx}

\usepackage{multirow}%
\usepackage{amsmath,amssymb,amsfonts}%
\usepackage{amsthm}%
\usepackage{mathrsfs}%

\title{Active magneto-mechanical metamaterial with the wave transmission and Poisson's ratio controlled via the magnetic field}

\author[1,2]{Krzysztof K. Dudek}
\author[3]{J. A. Iglesias Mart\'{i}nez}
\author[4]{L. Hirsinger}
\author[4]{M. Kadic}
\author[2]{M. Devel}

\affil[1]{Institute of Physics, University of Zielona Gora, ul. Szafrana 4a, Zielona, Gora 65-069, Poland}

\affil[2]{SUPMICROTECH, Universit\'{e} de Franche-Comt\'{e}, CNRS, Institut FEMTO-ST, F-25000 Besançon, France}

\affil[3]{Institut Jean Lamour, CNRS UMR 7198, University Lorraine, 54011 Nancy Cedex, France}

\affil[4]{Universit\'{e} de Franche-Comt\'{e}, CNRS, Institut FEMTO-ST, F-25000 Besançon, France}

\date{}

\begin{document}

\maketitle

\begin{abstract}
In recent years, there has been a notable increase in the significance of active mechanical metamaterials capable of being remotely manipulated through changes in external stimuli. While research in this area has achieved considerable success in controlling reconfiguration to induce shape morphing or alter static mechanical properties, the active control of wave propagation within these systems remains largely unexplored. In this study, we propose a magneto-mechanical metamaterial that can be entirely governed by adjusting the magnitude and direction of an external magnetic field. We demonstrate that such a system offers remote control over its Poisson's ratio, allowing for transitions from strongly auxetic to positive Poisson's ratio configurations. Additionally, our system enables manipulation of the phononic band structure, facilitating the formation of complete band gaps across different frequency ranges depending on the stage of the magnetically guided reconfiguration process. The potential for achieving such active control over the properties and behavior of these materials holds great promise for various applications, including robotics and smart vibration dampers that can be remotely controlled. 
\end{abstract}

\section{Introduction}

Mechanical metamaterials \cite{Evans_1991_Nature, WOJCIECHOWSKI198960, PhysRevLett.113.175503, craster2023mechanical, kadic20193d, Coulais_Smiley_Nature} are a class of structures capable of exhibiting counterintuitive mechanical properties thanks to their rational design. Over the years, such systems have been thoroughly investigated from the perspective of their ability to exhibit properties such as negative Poisson's ratio \cite{Grima_squares_2000, Bertoldi_Mullin_Adv_Mater_2010, Babaee_Bertoldi_Adv_Mater_2013, Ren_SMS_2016, Dudek_Adv_Mater_2022, REN2022108584, MIZZI2023110739, MIZZI2022111428, PhysRevApplied.7.024012, PhysRevE.108.045003}, negative stiffness \cite{TAN2019397, Tan_Kadic_Adv_Intell_Sys_2023, https://doi.org/10.1002/adma.201603959, DUDEK2020108403}, or high energy absorption \cite{https://doi.org/10.1002/admt.201800419, IMBALZANO2016339, Duncan_SMS_2016, NOVAK2022116174, CHEN2020105288, DUNCAN2023104922} that have proven to be of great significance in the case of applications ranging from biomedical \cite{ZADPOOR201941, Zadpoor_Mirzaali_2023} to soundproofing devices \cite{Cummer_Alu_Nature_2016, ZANGENEHNEJAD2019100031, 10.1121/1.3569707, https://doi.org/10.1002/admt.202100698, Airoldi_Ruzzene_2011}. Despite their numerous advantages, standard mechanical metamaterials typically share one limitation. Namely, once they are fabricated, it is very challenging to modify their properties. To do this, it is typically required to refabricate the system while implementing a different design. This, in turn, significantly reduces the applicability of such structures since they cannot be continuously used without sacrificing their efficiency. The solution to this problem turned out to be active mechanical metamaterials, i.e. structures that can undergo changes in their mechanical properties and deformation pattern based on the variation in the external stimulus. In recent years, it has been shown that it is possible to design metamaterials that can be controlled via temperature \cite{XIA20161, Pasini_Sci_Rep_2019, Tang_PNAS_2019, Ji_Kadic_Commun_Mater_2021, Korpas_2021}, magnetic field \cite{doi:10.1126/sciadv.aau6419, doi:10.1126/sciadv.abc6414, Heyderman_Nature_2019, GALEA_Compos_Struct_2022, Montgomery_2020, Dudek_MR_2013_magnets}, light \cite{https://doi.org/10.1002/adma.201906233}, and other stimuli \cite{Gladman_Lewis_Nat_Mater_2016}. This concept is very promising since it allows to achieve the fully programmable and remote control over the behavior of a metamaterial not only without the need of refabricating it but also without being in direct contact with the system. The latter is particularly important in the case of applications related to microrobotics \cite{Heyderman_Nature_2019, Wehner_Lewis_Nature_2016, Gazzola_Mahadevan_2014}, where due to a small scale it is not possible to influence the deformation of the structure through direct physical manipulation.

In the studies focused on active mechanical metamaterials it was shown that through the appropriate composite design, it is possible to almost arbitrarily adjust the response of the system depending on a change in the external stimuli. One of the main directions of studies related to stimuli-responsive systems is their shape-morphing ability \cite{Heyderman_Nature_2019, Gladman_Lewis_Nat_Mater_2016} which is very useful in the case of multiscale robotics. Active metamaterials can also change their static mechanical properties such as the Poisson's ratio \cite{GALEA_Compos_Struct_2022, Dudek_Kadic_Adv_Mater_2023, Galea_PSSB_2022} or stiffness \cite{Dudek_stiffness_PRSA_2018, DUDEK_magnets_Mater_Des_2020}. In fact, in recent studies \cite{GALEA_Compos_Struct_2022, Dudek_Kadic_Adv_Mater_2023, Galea_PSSB_2022}, it has been demonstrated that based on the changes in the external magnetic field, metamaterials can undergo a remotely controlled transition from a strongly auxetic to a nonauxetic configuration. This offers a considerable advantage over many standard auxetic metamaterials, where such a change in the properties of the system requires its refabrication. Another important avenue pursued by researchers working on active mechanical metamaterials is the possibility of controlling the phononic band structure and the transmission of waves through the system. In fact, over the years, there have been multiple studies reported, where it was shown that through the application of the external stimulus such as a change in the external magnetic field, it is possible to change the phononic band structure of the system \cite{Pierce_Matlack_SMS_2020, Moghaddaszadeh_Commun_Mater_2023, ZHANG2023101957, ZHANG2024112648, Sim_Renee_Adv_Mater_2023, Mohammadi_J_Appl_Mech_2019, KUNIN2016103}. This goal had been achieved amongst others through the use of soft magnetoactive laminates \cite{Sim_Renee_Adv_Mater_2023, Mohammadi_J_Appl_Mech_2019} and structures, where internal components exhibit the reorientation due to the presence of the external field \cite{ZHANG2023101957}. However, it is much more difficult to find studies reporting the possibility of achieving simultaneous control over wave transmission through the system as well as its Poisson's ratio. Notably, such attempts have been made but usually, the high level of control over one aspect of the behavior of the system comes at the expense of another \cite{Dudek_Kadic_Adv_Mater_2023, KUNIN2016103}. It becomes even more challenging when the mechanical reconfiguration caused by the application of the external magnetic field is expected to cause a large reconfiguration leading to a very considerable change in the volume of the system. This, in turn, is essential to benefit from the controllable Poisson's ratio in the case of many applications in engineering. Hence, the main objective of this work is to address these challenges. 

In this work, we propose a novel magneto-mechanical active metamaterial having its reconfiguration process controlled via the external magnetic field. We show that through changes in the orientation of such stimulus, the structure can follow very different deformation patterns that also correspond to very different values of Poisson's ratio with the concurrent large changes in its volume. In addition to changes in the static mechanical properties, we also show that programmable changes in the configuration of the system can at the same time lead to significant differences in the phononic band structure including the appearance of complete phononic band gaps.

\section{Concept}

The main objective of this work is to construct a magneto-mechanical metamaterial capable of exhibiting a versatile mechanical behavior based on the changes in the external magnetic field. To this aim, the behavior of the system should be controlled both by the magnitude and direction of the external field. More specifically, changes in the magnitude of the external stimulus are expected to guide the reconfiguration process so that the structure can follow a specific deformation pattern. On the other hand, a change in the direction of the external field is expected to result in a new deformation profile that is different from the ones observed for other spatial orientations of such stimulus. The latter is of particular significance since the possibility of following distinct deformation profiles is essential for the structure to exhibit different mechanical behavior.

\begin{figure}[!h]
	\centering
	\includegraphics[width=0.95\textwidth]{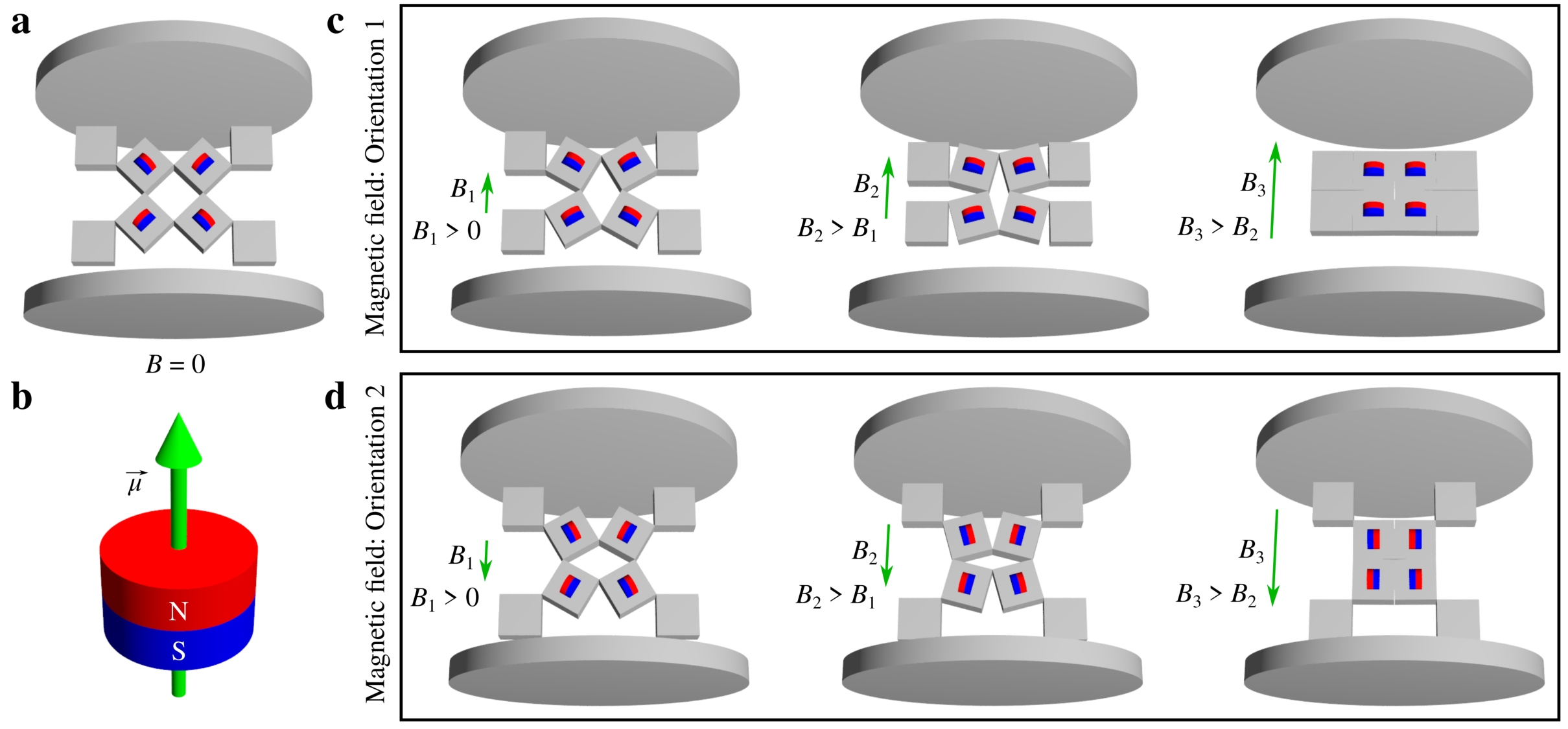}
	\caption{The concept of the remotely-controlled deformation process of the magneto-mechanical metamaterial subjected to magnetic fields having opposite orientations. {\bf a} The initial undeformed structure. {\bf b} A diagram showing an auxiliary magnet and the schematic representation of its magnetic dipole moment $\vec{\mu}$. The reconfiguration of the system originates from such dipole moments attempting to align with the external magnetic field. {\bf c} Deformation of a system subjected to the external magnetic field having orientation 1. {\bf d} The reconfiguration of the structure subjected to the magnetic field having orientation 2.}
	\label{fig:1}
\end{figure}

Objectives described above can be achieved through the use of the model schematically portrayed in Fig. \ref{fig:1}. Here, the considered metamaterial assumes the form of a nonmagnetic elastic lattice with small permanent magnets embedded in some of its structural units (see Fig. \ref{fig:1}b). Such a system is then placed within the uniform magnetic field induced by two poles of a large electromagnet. Since the field is uniform, whenever it has a non-zero magnitude, the magnets experience the magnetic torque that helps to align them with the orientation of the external field. Due to the specific design of the nonmagnetic lattice, the rotation of magnets embedded within the structure results in the reconfiguration of the entire metamaterial, where the extent of the mechanical deformation depends on the magnitude of the field.

As shown in Fig. \ref{fig:1}c-d, to assess the possibility of following different deformation profiles, in this work, we consider two opposite orientations of the magnetic field applied to the sample having the same initial position. For the first orientation of the field (see Fig. \ref{fig:1}c), the rotation of structural elements incorporating magnetic inclusions leads to the configuration, where the system resembles a solid block. On the other hand, for the opposite orientation (see Fig. \ref{fig:1}d), the deformation process results in the configuration resembling a chessboard, i.e. a square lattice composed of alternating solid blocks and voids of the same size. These clearly different deformation profiles show that the two orientations of the external magnetic field may affect the reconfiguration process and may potentially lead to a transition in the mechanical properties.   

\section{Model}
\subsection{Geometry}
\begin{figure}
	\centering
	\includegraphics[width=0.95\textwidth]{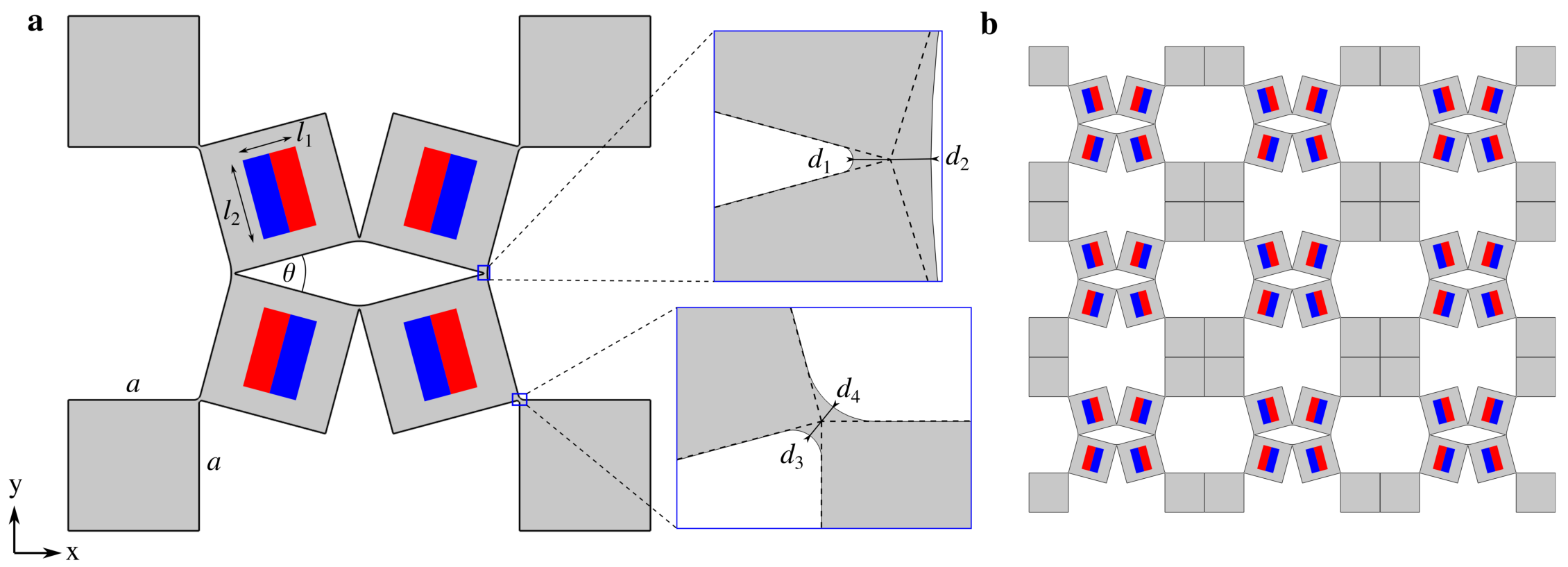}
	\caption{{\bf a} Cross-section of the unit cell of the considered system with insets portraying the way how hinges connecting the adjacent structural units were designed. {\bf b} An array consisting of $3 \times 3$ unit cells. The red and blue rectangles are used to indicate magnets and their specific polarity. Namely, the red and blue colors indicate the "north" and "south" poles respectively.}
	\label{fig:2}
\end{figure}

The metamaterial considered in this work is very similar to the conceptual diagrams portrayed in Fig. \ref{fig:1}. As shown in Fig. \ref{fig:2}a, the unit-cell of the system corresponds to the quasi-2D nonmagnetic lattice composed of square-like elements having a side length of $a$ = 7 mm that are connected through hinges ($d_{1} = d_{2} = d_{3} = d_{4} = 90 \, \mu$m). The four central structural units have magnetic inclusions in the form of cylindrical neodymium magnets embedded within the elastic material. Dimensions of these magnets in the $xy$ plane correspond to $l_{1} \times l_{2}$ = 3 mm $\times$ 5 mm, where quantities $l_{1}$ and $l_{2} $ can be described as the height and diameter of the cylindrical magnets (the 3D version showing the placement of magnets is provided in Supplementary Information). Finally, the angle between structural units with magnetic inclusions is defined as $\theta$. This parameter is of great significance since it quantifies the extent of the mechanical deformation of the system. Finally, it should be noted that to experimentally analyze the behavior of the considered metamaterial, a structure composed of $3 \times 3$ unit cells was considered. The out-of-plane thickness of such a system is equal to 7 mm.  

\subsection{Fabrication}

The nonmagnetic lattice was additively manufactured through the use of the Formlabs Form 3 3D printer. The resin used during the printing process was characterized to have Young's modulus of 3.2 MPa, Poisson's ratio of 0.4, and a density equal to 1200 kg m$^{-3}$. It was the standard Formlabs resin called "Elastic resin 50A" which is compatible with the Formlabs Form 3 3D printer. Magnets embedded within the lattice were commercial N38 neodymium magnets with residual induction around 1200 mT. 

\subsection{Experiments}

To analyze the evolution of the structure, it was placed within the approximately uniform magnetic field induced by the electromagnet. Then its behavior was analyzed for two orientations of the magnetic field as shown in Fig. \ref{fig:1}. In the first experiment corresponding to the Orientation 1 of the magnetic field, the system was subjected to the magnetic field corresponding to a gradually increasing magnetic flux density ranging from 0 mT to 70 mT with the step of 5 mT. For every value of the magnetic flux density, the system was allowed to assume the equilibrium configuration within 60 seconds after which time the magnitude of the field was changed. After reaching the final fully locked configuration at 70 mT, this process was repeated at the range between 70 mT and 0 mT to show the reversibility of the considered remotely controlled reconfiguration.

The experiment conducted for the second orientation of the magnetic field (see Fig. \ref{fig:1}) was very similar. However, in this case, the maximum field required to reach the fully deformed configuration was significantly smaller and the magnetic flux density assumed the value of 35 mT.

\subsection{Poisson's ratio} \label{Model_Poisson}

The considered model is designed in a way, where all structural blocks are connected through very thin hinges. As a result, during the mechanical deformation, the square-like structural units can retain their shape. In such a case, if we assume that the structure follows a symmetric deformation process, changes in its linear dimensions would depend solely on one parameter, i.e. angle $\theta$. In this case, the dimensions of the unit cell of the considered system in the $xy$ plane can be computed as follows:

\begin{equation}
L_{x} = 2a \left[\cos \left(\frac{\theta}{2} \right) + \sin \left(\frac{\theta}{2} \right) + 1 \right] \quad \quad \textrm{and} \quad \quad L_{y} = 2a \left[1 +  \cos \left(\frac{\theta}{2} \right) \right] .
\end{equation} 

Analytical expressions defining linear dimensions can be later used to calculate Poisson's ratio of the system as shown below:

\begin{equation}
\nu_{yx, eng} = - \frac{\varepsilon_{x, \, eng}}{\varepsilon_{y, \, eng}} \quad \quad \textrm{and} \quad \quad \nu_{yx, inc} = - \frac{\varepsilon_{x, \, inc}}{\varepsilon_{y, \, inc}}
\label{Poisson}
\end{equation}

In Eq. \ref{Poisson}, $\varepsilon_{x, \, eng}$ and $\varepsilon_{y, \, eng}$ stand for the engineering strains in the $x$ and $y$ dimensions. Similarly, $\varepsilon_{x, \, inc}$ and $\varepsilon_{y, \, inc}$ are incremental strains. Given an example of the deformation in the $x$-direction, these strains can be defined as $\varepsilon_{x, \, eng} = \left (L_{x}(t) - L_{x}(t=0) \right) / L_{x}(t=0)$ and $\varepsilon_{x, \, inc} = \left(L_{x}(t+dt) - L_{x}(t) \right) / L_{x}(t)$. The difference between these two definitions is the fact that the engineering strain is calculated relative to the initial dimensions of the unit cell. On the other hand, the incremental strain depends on the change in the linear dimensions of the unit cell that is calculated based on the consecutive stages of the deformation process.

\section{Results and Discussion}

\subsection{Stimuli-responsive reconfiguration}

The main objective of this work is to show that through changes in the external magnetic field, it is possible to change both the Poisson's ratio and the phononic band structure of the considered active metamaterial. However, before attempting to reach these goals, it is necessary to demonstrate that the analyzed structure can undergo a significant remotely-controlled reconfiguration process since the changes in the aforementioned properties are expected to originate from the variation in the geometry of the system.

\begin{figure}
	\centering
	\includegraphics[width=0.95\textwidth]{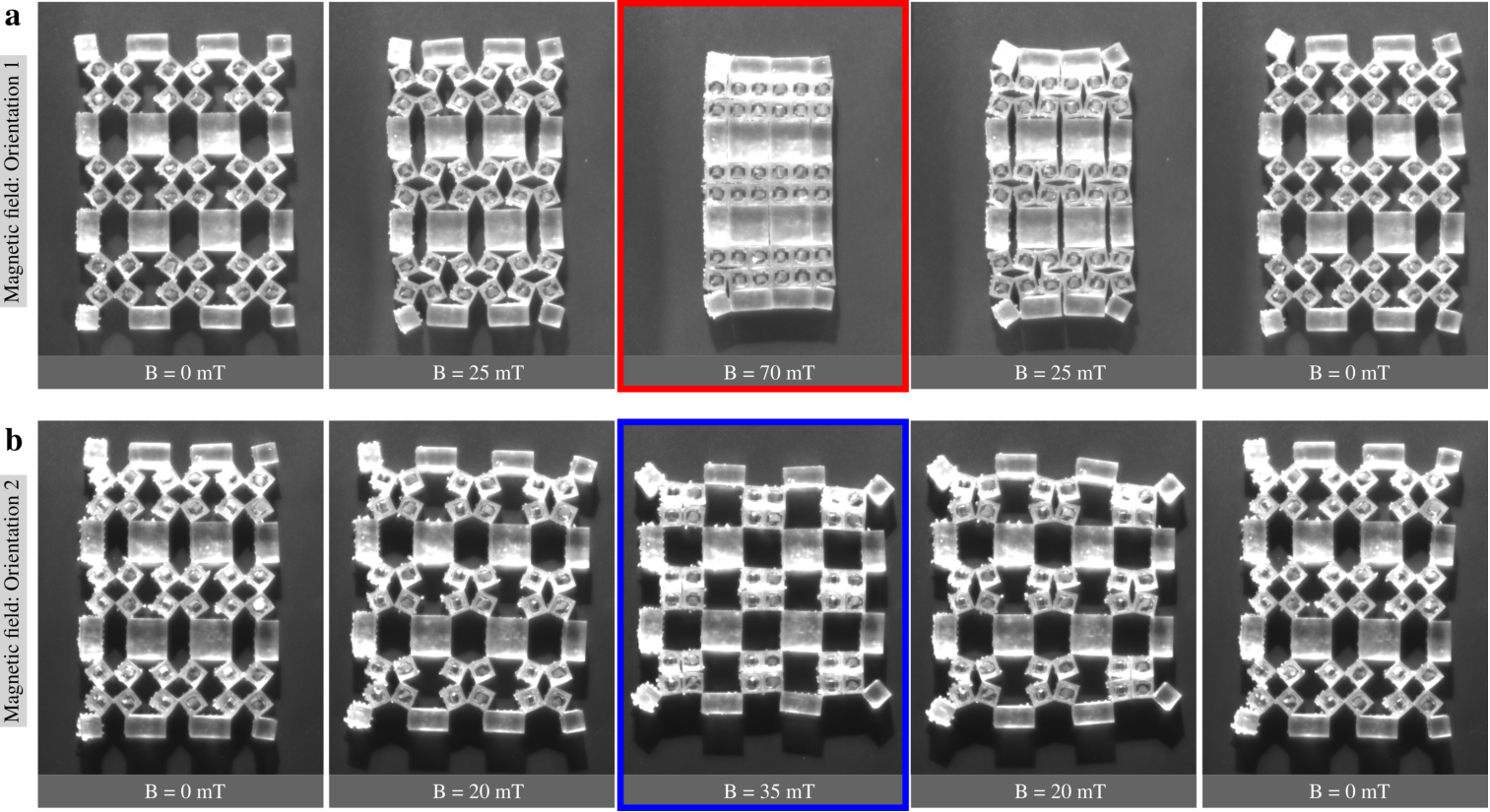}
	\caption{Deformation of the magneto-mechanical metamaterials guided by the changes in the magnitude of the external magnetic field. Here, we conduct two experiments, where the same sample is subjected to the external magnetic fields having opposite orientations that are referred to as {\bf a} Orientation 1 and {\bf b} Orientation 2. Blue and red frames surrounding one configuration for Orientations 1 and 2 of the field indicate configurations that are maximally deformed. In other words, the further increase in the magnetic flux density does not change these configurations since they are geometrically locked.}
	\label{fig:3}
\end{figure}

As shown in Fig. \ref{fig:3}a, once the metamaterial is subjected to the external field having Orientation 1 (see Fig. \ref{fig:1}), the angle $\theta$ tends to increase which leads to the situation where the areas of apertures formed between the adjacent structural units are gradually decreasing (see Supplementary Video 1). In fact, once the magnetic flux density related to the external field becomes relatively large and reaches the value of 70 mT, all apertures are completely closed and the entire structure assumes the configuration resembling a solid block (see the configuration highlighted by the red frame in Fig. \ref{fig:3}a). At this point, the structural units can no longer continue their rotation under the influence of the external field. Hence, the only possible type of further in-plane reconfiguration corresponds to the backward rotation of such elements. To explore this scenario, one can analyze what happens if after reaching the locked configuration, the magnetic flux density associated with the external field is gradually reduced to the original value of 0 mT. Based on the provided results, it is clear that such a change in the magnitude of the external stimulus leads to the reopening process which continues until the system assumes once more the configuration, where $\theta \approx 90^{\circ}$. This, in turn, indicates that the considered active reconfiguration process is reversible.

In the case of the sample subjected to the magnetic field having the orientation referred to as Orientation 1, it was already demonstrated that the active reconfiguration of the system can be induced and controlled via changes in the magnitude of the external stimulus. Nonetheless, it is also interesting to analyze what happens when the direction of such a field changes relative to the sample. To this aim, the second experiment was conducted, where the direction of the magnetic field was reversed while the orientation of the sample was maintained (see Supplementary Video 2). As shown in Fig. \ref{fig:3}b, this change in the external field strongly influenced the deformation process and resulted in the opposite rotation of the structural units. Namely, as the magnetic flux density of the external field was increasing, the angle $\theta$ was decreasing as opposed to what was the case for Orientation 1 of the field. Such reconfiguration was continued until the point where the magnetic flux density reached 35 mT and structural units with embedded magnets got into direct contact with each other. However, it is important to note that in this scenario, the final configuration of the system was very different from the one observed before (see the blue frame in Fig. \ref{fig:3}b), i.e. some of the apertures were not closed as a result of the reconfiguration. Instead, the entire system assumed the configuration resembling a chessboard with solid blocks associated with the sample and square-shaped voids corresponding to apertures. Furthermore, based on Fig. \ref{fig:3}b, it is possible to observe that also for this orientation of the magnetic field the deformation process is reversible, and once the magnetic flux density assumes the value of 0 mT, the structure returns approximately to its initial orientation. 

At this point, it is important to note that for the magnetic fields having both orientations 1 and 2, the angle $\theta$ changed by $90^{\circ}$ to reach the locked configuration (from $90^{\circ}$ to $180^{\circ}$ in the first configuration and from $90^{\circ}$ to $0^{\circ}$ in the second configuration). However, the magnetic flux density of the applied field that was needed to do it, in one case, was significantly larger than for the other orientation. Namely, for Orientation 1 the required magnetic flux density of the external field assumed the value of 70 mT, while for Orientation 2 it was 35 mT. This large difference can be explained based on Fig. \ref{fig:1}. From this diagram, it is clear that close to the end of the deformation process, the resultant magnetic dipole moments of the magnets were almost parallel to the external field. This means that the magnetic torques acting on the magnets were weak and it was necessary to apply a relatively large field to complete the reconfiguration process. In contrast, at the same stage of the deformation process, magnetic dipole moments of magnets subjected to the field having orientation 2 were almost perpendicular to its lines which resulted in relatively large magnetic torques. Thus, a considerably smaller magnetic flux density of the external field was needed to complete the reconfiguration in this case. It should also be noted that in addition to the interaction of magnets with the external magnetic field, the magnets were also interacting with each other. However, although the extent to which the system was deforming was partly dependent on internal interactions, these interactions were not strong enough to cause any visible reconfiguration on their own due to the separation distance and relative orientations. Thus, the external field was the dominant factor responsible for the deformation process.

\subsection{Remote control of the Poisson's ratio}

Based on the conducted experiments and the results presented in Fig. \ref{fig:3}, it is clear that depending on the orientation of the external magnetic field, the magneto-mechanical metamaterial can follow very different deformation patterns. Here, the question arises whether such different reconfigurations can influence static mechanical properties of the system such as its Poisson's ratio. To this aim, one can use the equations proposed in section 3.4. 

\begin{figure}
	\centering
	\includegraphics[width=0.99\textwidth]{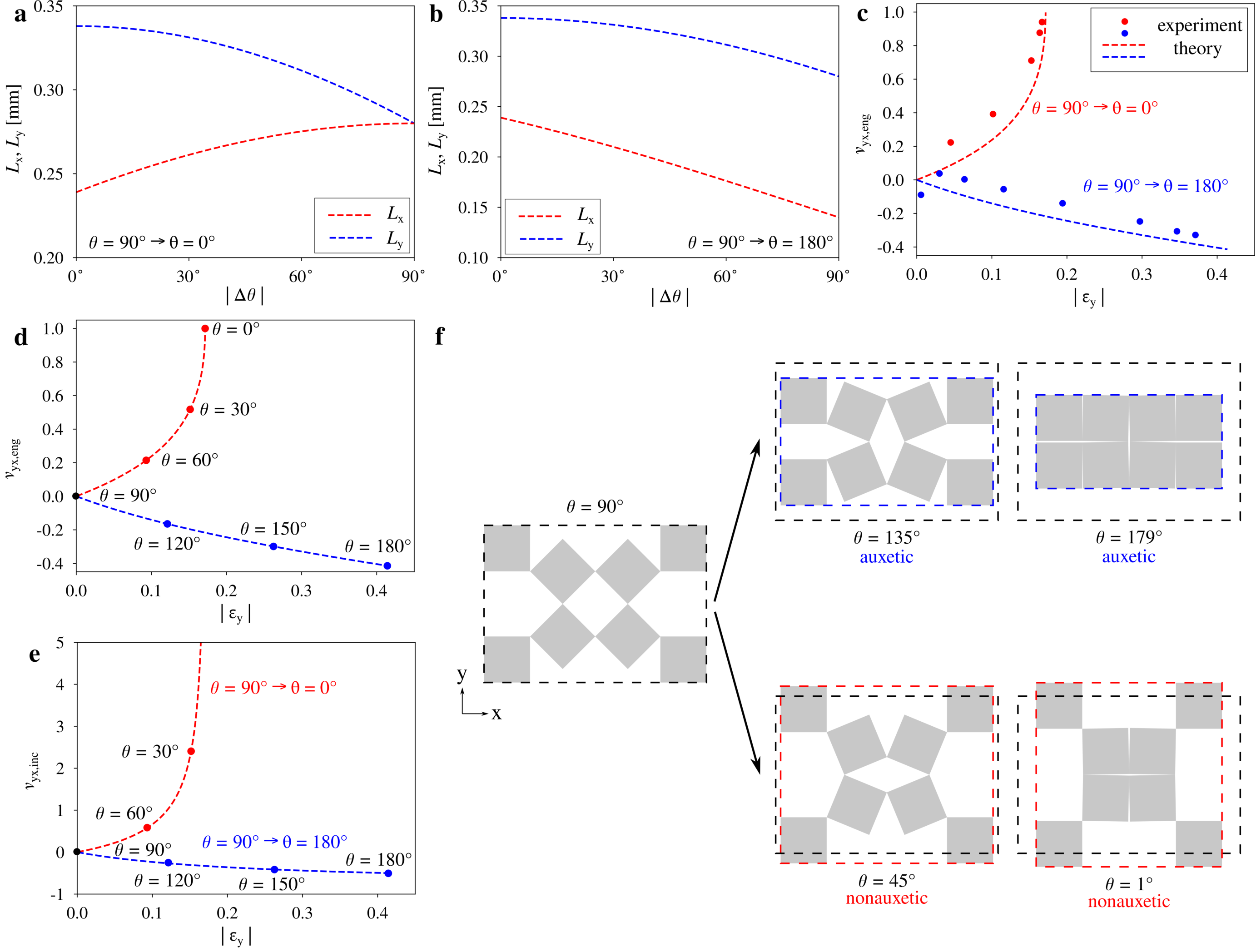}
	\caption{{\bf a} Changes in the linear dimensions of the unit cell throughout the deformation process corresponding to a change from $\theta = 90^{\circ}$ to $\theta = 0^{\circ}$. {\bf b} Variation in linear dimensions of the unit cell during the reconfiguration associated with a change in $\theta$ from $90^{\circ}$ to $180^{\circ}$. {\bf c} Comparison of the experimental and theoretical results related to the engineering Poisson's ratio. {\bf d} Theoretical engineering Poisson's ratio measured along the $y$-axis (direction of the magnetic field) for the two considered orientations of the magnetic field. {\bf e} Theoretical incremental Poisson's ratio (see Section \ref{Model_Poisson}, where quantity $\nu_{yx,inc}$ is defined in Eq. \ref{Poisson}). Both on panels d and e, additional dots indicate specific values of $\theta$ that correspond to the structure at different stages of the deformation process. {\bf f} The diagram shows a difference in the deformation patterns of the sample subjected to the magnetic field having two different orientations.}
	\label{fig:4}
\end{figure}

According to Fig. \ref{fig:4}c-d, the structure subjected to the magnetic field having orientation 1, where $\theta$ changes from $90^{\circ}$ to $180^{\circ}$, exhibits a strong auxetic behavior that reaches the value around -0.4. On the other hand, the change in the orientation of the field ($\theta = 90^{\circ} \rightarrow \theta = 0^{\circ}$) leads to a highly positive Poisson's ratio at large strains. This behavior can be observed both in the case of theoretical model and experimental analysis with both approaches being in a good agreement. Furthermore, similar trends can also be observed for the incremental Poisson's ratio as shown in Fig. \ref{fig:4}e. This means that based solely on the variation in the direction of the magnetic field, the considered system can undergo a remotely induced transition from the auxetic to the nonauxetic configuration which is not very common in the field of mechanical metamaterials. To better understand the difference in the behavior of the metamaterial when subjected to the field having orientation either 1 or 2, one can analyze what happens when the system is subjected to the reconfiguration process corresponding to either the change in $\theta$ from $90^{\circ}$ to $0^{\circ}$ or from $90^{\circ}$ to $180^{\circ}$. As shown in Fig. \ref{fig:4}a, as $\theta$ changes from $90^{\circ}$ to $0^{\circ}$, Poisson's ratio assumes positive values and grows very rapidly with the increase in strain. This stems from the fact that the system is strongly anisotropic and during this specific deformation process, the $L_{x}$ dimension of the unit cell gradually increases while the $L_{y}$ dimension is decreasing at the faster rate. As a result, when the value of $\theta$ approaches $0^{\circ}$, the rate of growth in $L_{x}$ is very small while $L_{y}$ decreases very fast. On the other hand, as shown in Fig. \ref{fig:4}b, during the reconfiguration corresponding to a change in $\theta$ from $90^{\circ}$ to $180^{\circ}$, both of the linear dimensions gradually decrease which leads to a gradual decrease in the value of Poisson's ratio. It should be also noted that the origin of the change in the Poisson's ratio profile for the two types of deformation can be explicitly deduced from the diagrams portraying different stages of the reconfiguration (see Fig. \ref{fig:4}f). Finally, it should be noted that as indicated in Fig. \ref{fig:4}d-e by means of auxiliary markers, Poisson's ratio changes nonlinearly with the increase / decrease in $\theta$ which behavior stems from the nonlinear changes in the dimensions of the unit cell.

\subsection{Active control over the phononic band structure of the system} \label{section_active}

The primary objective of this work is to show the possibility of the remote control over the wave propagation within the considered magneto-mechanical metamaterial. In addition to the possibility of actively controlling Poisson's ratio, this ability opens many opportunities from the perspective of the multifunctional applications used in robotics or medicine. Furthermore, as mentioned in the Introduction, the possibility of significantly changing the phononic band structure of the system that leads to changes in the wave propagation is very rare and typically requires reconstructing the system or at least changing some of its features such as the thickness of its hinges \cite{https://doi.org/10.1002/pssb.202200404}. However, in this work, we will attempt to show that this goal can be achieved solely as a result of the reconfiguration that can be remotely controlled as already presented in the previous section (see section 4.1).

\begin{figure}
	\centering
	\includegraphics[width=0.99\textwidth]{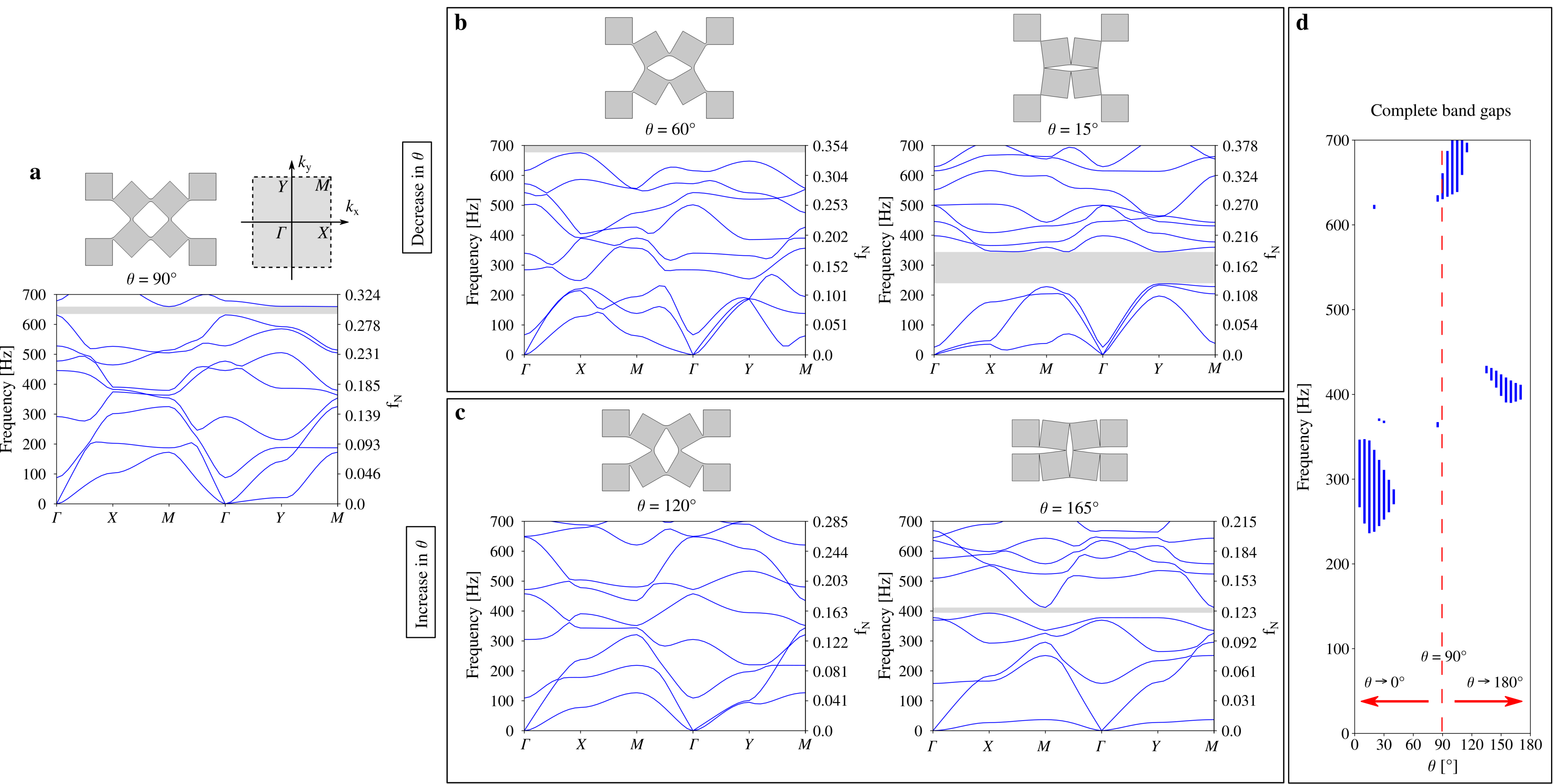}
	\caption{{\bf a} The elementary unit cell corresponding to the initial configuration in real space, where $\theta = 90^{\circ}$, as well as the schematic diagram of the first Brillouin zone. The bottom part of the panel shows the phonon dispersion corresponding to this configuration. {\bf b} Changes in the phononic band structure upon decreasing the initial value of $\theta = 90^{\circ}$. {\bf c} Phonon dispersion associated with configurations, where $\theta > 90^{\circ}$. On panels b and c, the complete band gaps are indicated by means of the grey rectangles corresponding to specific ranges of frequencies. {\bf d} Complete band gaps for different configurations of the system. In addition to the frequency, all graphs include the normalized frequency $f_{N}$ which was calculated as $f_{N} = f\frac{L_{y}}{\sqrt{E / \rho}}$, where the respective variables stand for: $f$ - frequency, $E$ - Young's modulus, $\rho$ - density, $L_{y}$ - linear dimension of the unit cell.}
	\label{fig:5}
\end{figure}

In this work, the phonon dispersion was determined for the unit cell corresponding to different 2D configurations that this system assumes throughout the deformation process. To this aim, Floquet boundary conditions were implemented in the $x$ and $y$ direction of the rectangular cell shown in Fig. \ref{fig:5}a. All calculations were conducted through the use of the COMSOL Multiphysics software that operates based on the Finite Element Method.

The starting point for our considerations is the phononic band structure associated with the configuration, where $\theta = 90^{\circ}$. Our goal will be to show what happens to the phonon dispersion once this geometry evolves to configurations with $\theta$ either in the range between $90^{\circ}$ and $180^{\circ}$ or $0^{\circ}$ and $90^{\circ}$ similarly to the reconfiguration processes observed in the experiment. According to Fig. \ref{fig:5}a, the phononic band structure for the reference system associated with $\theta = 90^{\circ}$ does not have any complete band gaps at the low frequencies related to the first few bands. Indeed, frequency must reach 632~Hz (top of the 8th band) to see the bottom of the first complete band gap which extends up to 659 Hz (bottom of the 9th band). Of course, one can note some directional bands at lower frequencies. However, here, we are interested in the possibility of blocking the potential propagation of waves in all planar directions of the considered quasi-2D model. Hence, we are paying close attention to complete band gaps. 

As shown in Fig. \ref{fig:5}b, as the configuration of the system changes and $\theta$ becomes smaller than $90^{\circ}$, it is evident that the phononic band structure significantly deviates from its form associated with the initial configuration. It is also possible to see that as $\theta$ decreases to relatively low values such a broad complete band gap can be observed. This band gap assumes its maximum size for $\theta \approx 15^{\circ}$, where it corresponds to the range of frequencies between 238 Hz and 344 Hz (see Fig. \ref{fig:5}d). This change in the phononic band structure means that solely by inducing the deformation of the considered metamaterial corresponding to the decrease in $\theta$, it is possible to transition from the configuration, where waves are freely transmitted through the system to the configuration, where this is not possible at a given range of frequencies. A similar analysis can also be conducted for the deformation process, where the value of $\theta$ is gradually increasing. As shown in Fig. \ref{fig:5}c, in this case, however, the phononic band structure changes in a very different manner. Most notably, for a vast part of the deformation process, it is hard to observe any complete band gap. Only when reaching strongly deformed configurations associated with the value of $\theta$ exceeding $140^{\circ}$, it is possible to observe some complete band gaps that are not negligible. Nonetheless, these band gaps corresponds to higher frequencies than those observed for the analogical process associated with the decrease in $\theta$. The largest complete band gap observed in this case can be seen for $\theta \approx 155^{\circ}$, where it corresponds to the range of frequencies between 392 Hz and 419 Hz (see Fig. \ref{fig:5}d). Finally, it should be highlighted that the possibility of controlling the wave transmission through the system by changing its configuration was also verified experimentally (see Fig. S1). As discussed in the Supplementary Information, similarly to the theoretical predictions, considerable complete band gap was observed at relatively low frequencies for the configuration associated with $\theta = 15^{\circ}$. On the other hand, for the configuration corresponding to $\theta = 90^{\circ}$, there were no complete gaps in the transmission at the considered range of frequencies.

\subsection{Experimental verification of the possibility of changing the transmission through the system via its reconfiguration}

In the previous section (see Section \ref{section_active}), it has been demonstrated that a change in the configuration of the system can lead to a change in its phononic band structure. Most importantly, it was shown that such reconfiguration can lead to a significant variation in the phononic band gaps corresponding to the system. However, while the possibility of observing the active reconfiguration was assessed experimentally, the ability to observe the variation in the phononic band gaps was examined solely from the perspective of the FEM simulations. This stems from the fact that to experimentally demonstrate the active reconfiguration, a very soft resin was used to fabricate the system which makes the analysis of the wave propagation to be very difficult. Nevertheless, this does not mean that the concept of the change in the transmission through the system as a result of its reconfiguration cannot be validated experimentally. To do it, one can fabricate individual samples associated with specific configurations of the system through the use of stiff resin that would make transmission testing much more feasible and reliable. To this aim, we decided to pick two specific configurations of our model that correspond to $\theta = 15^{\circ}$ and $\theta = 90^{\circ}$. The selection of these two configurations was not coincidental since as shown in Fig. \ref{fig:5}, for one of them ($\theta = 15^{\circ}$), one should expect to observe a gap in transmission at relatively low frequencies while for the other configuration, no such effect is expected. To fabricate the samples, a stiff resin (commercial ABS-like resin from the Anycubic company) corresponding to Young's modulus of 1.15 GPa, density $\rho = 1200$ kg m$^{-3}$ and $\nu = 0.4$ was used. In addition, to achieve reliable results, each sample was composed of $6 \times 6$ unit cells and their out-of-plane thickness was very considerable and reached 5 cm. To print such a large system, the linear dimension $a$ that describes the size of square-like elements was set to be equal to 3 mm while parameters corresponding to hinges were assumed to be the following $d_{1} = d_{2} = d_{3} = d_{4} = 120 \, \mu$m. All of the remaining details regarding the set up as well as the experiment are provided in the Supplementary Information.  

As shown in Fig. \ref{fig:6}a-b, before conducting the experimental analysis of the wave transmission, the behavior of the system was assessed through FEM simulations (the way how the samples were excited as well as the information about the boundary conditions is described in the Supplementary Information). Similarly to the predictions related to the phonon dispersion (see Fig. \ref{fig:6}a), at the range of frequencies between 12531 Hz and 17440 Hz, one can observe a complete phononic band gap that matches the transmission gap observed in the simulations (see Fig. \ref{fig:6}c). On the other hand, no such gap was observed for the configuration, where $\theta = 90^{\circ}$. Furthermore, as shown in Fig. \ref{fig:6}c and Fig. \ref{fig:6}e, upon comparing the computational and experimental transmission results associated with the configuration of $\theta = 15^{\circ}$, it is clear that the two sets of results are in a good agreement and that the predicted transmission gap can be also observed in the experiment. Of course, here, in the experiment, we consider only two specific directions that can be conveniently assessed in the case of the considered rectangle-shape system (see Fig. \ref{fig:6}b-c). Furthermore, it should be noted that for $\theta = 90^{\circ}$, no transmission gap can be observed in both of these directions. These results confirm the theoretical predictions made in this work and the possibility of changing the transmission through the system as a result of its reconfiguration.           

At this point, one can also note that the amplitude of the signal during the transmission testing behaves in a very specific manner. Namely, upon increasing the frequency, one can observe an overall decrease in signal which can be primarily attributed to the inherent material losses, which are particularly significant in the polymer used in 3D printing technology. Additionally, since the structure is composed of a finite number of cells, the attenuation caused by the gaps is also limited. This means that while the signal at the majority of considered frequencies does experience some reduction, it is not sufficient to reach the noise floor. The exception from this trend are ranges of frequencies that can be interpreted as transmission gaps. In fact, as shown in Fig. \ref{fig:6}e, where we recognize the appearance of the transmission gap, this effect can be observed in both propagation directions at the aforementioned range of frequencies around 15 kHz. However, due to the increased material losses, at higher frequencies, the signal fails to recover fully to its original level. This highlights the frequency-dependent nature of the losses, which prevent the system from achieving complete signal restoration at higher-frequency ranges.

\begin{figure}
	\centering
	\includegraphics[width=0.99\textwidth]{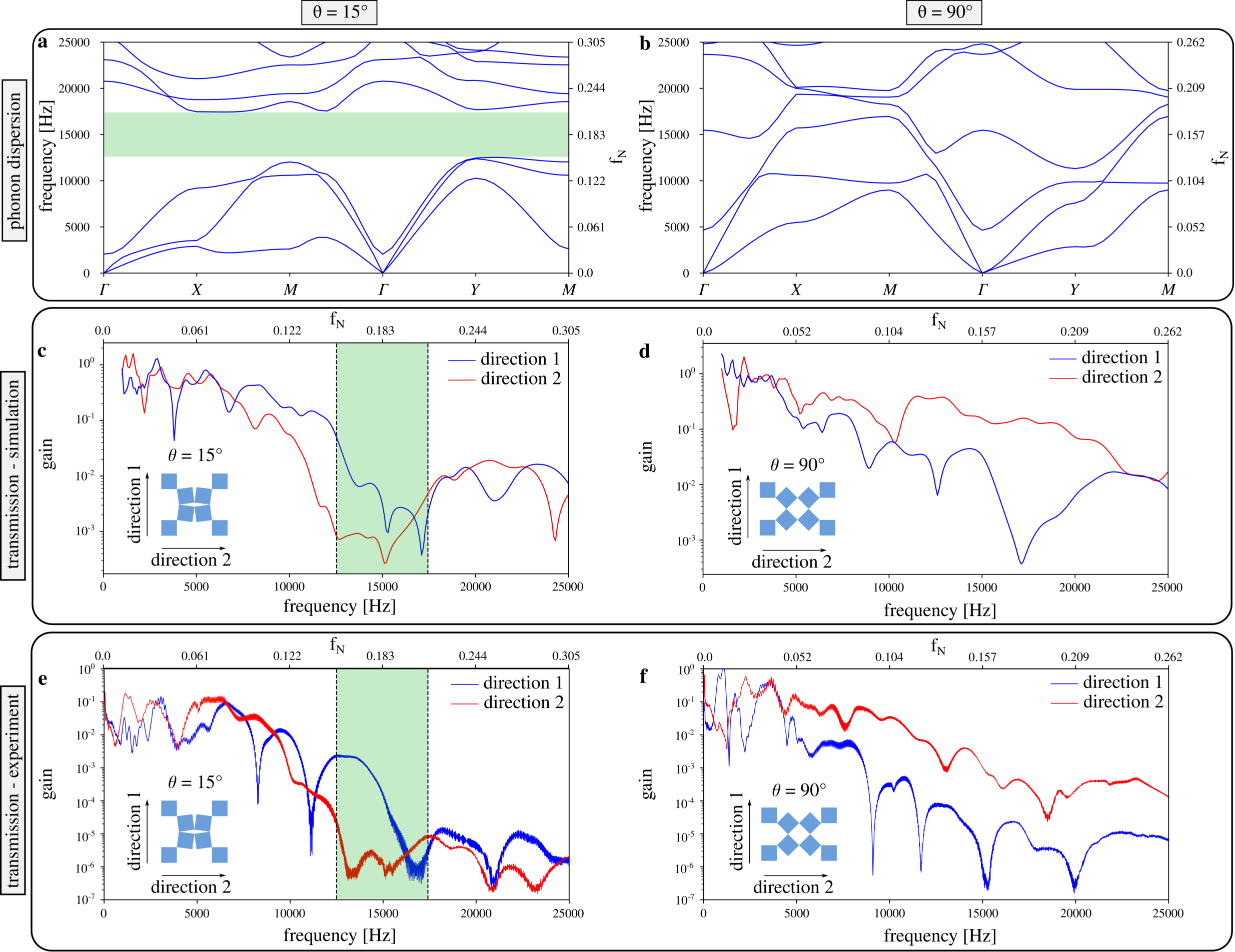}
	\caption{Comparison of the computational and experimental results corresponding to samples associated with $\theta = 15^{\circ}$ and $\theta = 90^{\circ}$ respectively. {\bf a-b} FEM results corresponding to the phonon dispersion for the considered configurations of the system. In this case, the green region indicates the complete phononic band gap. {\bf c-d} FEM results corresponding to the transmission through the finite system composed of $6 \times 6$ unit cells. {\bf e-f} Experimental results demonstrating transmission through the two samples considered samples. The quantity "gain" on the vertical axis of panels c-e stands for $|A_{out} / A_{in}|$, where $A_{in}$ is the amplitude of the signal close to the point that was used to excite the sample (see Supplementary Information). On the other hand, $A_{out}$ is the amplitude of the signal on the opposite side of the system. In addition to the frequency, all graphs include the normalized frequency $f_{N}$ which was calculated as $f_{N} = f\frac{L_{y}}{\sqrt{E / \rho}}$, where the respective variables stand for: $f$ - frequency, $E$ - Young's modulus, $\rho$ - density, $L_{y}$ - linear dimension of the unit cell.} 
	\label{fig:6}
\end{figure}

All of the results presented in this work are of great significance since it is shown that the considered magneto-mechanical metamaterial exhibits a fully programmable behavior, where through changes in the magnitude and direction of the external magnetic field it is possible to guide and control its reconfiguration process. It is shown that such control over the deformation is beneficial not only from the perspective of changing the geometry of the system but can also be used to alter its properties. More specifically, it allows for modification of both its Poisson's ratio and the phononic band structure. The possibility of remotely controlling Poisson's ratio offers many practical advantages ranging from the multi-scale shape morphing devices to changes in the energy absorption ability. Particularly in the case of shape morphing, the potential applicability of the proposed concept stems from the fact that depending on the orientation of the external field, the considered system can follow a different reconfiguration pattern and exhibit different values of Poisson's ratio. In the literature \cite{Dudek_Adv_Mater_2022}, it has been demonstrated that a graded metamaterial composed of unit cells being seemingly the same but associated with different values of Poisson's ratio can undergo very complex shape morphing solely as a result of a mismatch in this property. In the case of the considered metamaterial, one can achieve a similar effect. This can be done by changing the initial orientation of magnetic inclusions in different unit cells within the system. This way, upon subjecting the entire sample to the external magnetic field, for the sufficiently large number of unit cells, this metamaterial could undergo shape morphing into almost an arbitrary predefined pattern and come back to the initial state once the external field is switched off (see Fig. \ref{fig:3}). Furthermore, even though the above example is related to the metamaterial designed in a graded manner, there are also many practical advantages related to the uniformly designed versions of this system that do not have to rely on its Poisson's ratio. The tunable phononic band structure allows the construction of devices, where waves corresponding to a specific range of frequencies could be on-demand transmitted through the sample or damped. This is particularly interesting in terms of the smart insulators that could significantly change their mechanical performance without the necessity of being replaced. The only disadvantage of the current model is the fact that it incorporates macroscopic permanent magnets as magnetic inclusions that drive the mechanical reconfiguration. This approach, despite being very effective for macroscopic systems, is problematic to implement at much lower scales such as the microscale, where similar solutions would be very promising from the point of view of modern micro-robotics. However, one can hope that in future studies, one could attempt to propose an idea for the actuation mechanism that could replace the magnets. This hope seems to be quite feasible since in the literature, one can find multiple examples of complex reconfigurable micro-scale metamaterials that can be actuated by different stimuli and are suitable for use in microrobotics \cite{Ji_Kadic_Commun_Mater_2021, Heyderman_Nature_2019, Na_Hayward_Adv_Mater_2015}. One such potential idea could be related to the use of multimaterial 3D printing involving the use of a resin with a high thermal expansion coefficient in the vicinity of the hinges. In this case, the reconfiguration could be induced by the change in temperature.

%On the other hand, the tunable phononic band structure allows the construction of devices, where waves corresponding to a specific range of frequencies could be on-demand transmitted through the sample or damped. This is particularly interesting in terms of the smart insulators that could significantly change their mechanical performance without the necessity of being replaced. The only disadvantage of the current model is the fact that it incorporates macroscopic permanent magnets as magnetic inclusions that drive the mechanical reconfiguration. This approach, despite being very effective for macroscopic systems, is problematic to implement at much lower scales such as the microscale, where similar solutions would be very promising from the point of view of modern micro-robotics. Nonetheless, it is possible to fabricate miniaturized versions of the considered model that for example incorporate appropriately distributed magnetic nanoparticles. However, in this case, it is likely that a very strong magnetic field would be required to induce the active reconfiguration. Furthermore, the initial positioning of the magnetic nanoparticles with the periodic optimal initial orientation of their magnetic dipolar moments might be tricky. 

\section{Conclusions}
In this work, we showed that the proposed magneto-mechanical metamaterial can undergo a significant reconfiguration process depending on the external magnetic field. It was demonstrated that while changes in the magnitude of the magnetic field allow the structure to follow a specific deformation pattern, the variation in its direction makes it possible to completely modify the type of the reconfiguration process. Such possibility of controlling the deformation allows us to also achieve different values of Poisson's ratio depending on the variation in the external stimulus. More specifically, the considered metamaterial can exhibit either positive Poisson's ratio or strongly auxetic behavior depending solely on changes in the orientation of the magnetic field. In addition, at the same time, variation in the external field makes it possible to modify the phononic band structure corresponding to the system. Namely, it was shown that the active reconfiguration may cause the appearance of significant complete phononic band gaps at specific ranges of frequencies. All of these results indicate that the proposed metamaterial could be implemented in the case of devices utilizing both the possibility of controlling the Poisson's ratio and the wave propagation within the structure. One of the promising applications could be for example tunable smart dampers that could adjust their performance based on the changes in the external magnetic field.

\section*{Acknowledgements}
K.K.D. acknowledges the support of the Polish National Science Centre (NCN) in the form of the grant awarded as a part of the SONATA 18 program, project No. 2022/47/D/ST5/00280 under the name "Active micro-scale mechanical metamaterials controlled by the external stimuli". \\
This work was partially supported by a program of the Polish Ministry of Science under the title ‘Regional Excellence Initiative’, project no. RID/SP/0050/2024/1.
\\
K.K.D. thanks  SUPMICROTECH-ENSMM for a 2 months invited professor position and K.K.D and M.D. thank SUPMICROTECH-ENSMM for a financial support through a BQR project.\\
This work has also been supported by the EIPHI Graduate School (contract ANR-17-EURE-0002).  \\
The authors acknowledge the support of the ANR PNanoBot project (contract "ANR-21-CE33-0015").



%\bibliographystyle{elsarticle-num} 
%\bibliography{bibliography_file_ver2.bib}


\end{document}